\documentstyle[12pt]{article}

\author{Ji\v{r}\'{\i} Ho\v{s}ek\\
Nuclear Physics Institute\\
Czech Academy of Sciences\\
25068 \v{R}e\v{z} near Prague, Czech Republic\\
and\\
Georges Ripka\\
Service de Physique Th\'eorique\\
Centre d'Etudes de Saclay\\
F-91191 Gif-sur-Yvette Cedex, France}
\title{Macroscopic Chromodynamics of a Confining Medium
}

\begin{document}

\maketitle
\begin{abstract}
A nonminimal interaction of the gluon field with a massive antisymmetric
tensor order parameter is shown to yield a vanishing dielectric function at
tree level. The London-type description of this perfect diaelectric suggests
a tractable non-Abelian chromodynamics of the Ginzburg-Landau type which is
suitable for describing the confinement-deconfinement phase transition.\\
\vskip 0.5cm

PACS\ Numbers: 12.38\ Aw, 11.15\ Kc, 12.38 Mh
\end{abstract}

We still lack a microscopic understanding, in terms of QCD, of the
experimental fact that colored quarks and gluons do not escape separately
from colorless hadrons. The confinement of color can however be understood
if the vacuum is assumed to exhibit the property of a dual Meissner effect,
that is, if it behaves as a perfect color diaelectric \cite{thirring},\cite
{Lee},\cite{Kogut},\cite{Chodos}. This letter presents a phenomenological
London-Ginzburg-Landau type model exhibiting {\it bona fide} this property.
The model provides for an operational framework suitable for the description
of various properties of the confinement-deconfinement phase transition at
finite temperature and in strong chromoelectric fields. We hope that our
phenomenological model may ultimately serve as a guide towards a more
microscopic description.

The basic question is: which order parameter can produce a gauge invariant
interaction with the gluon field $A_a^\mu $ such that the color dielectric
function $\epsilon $ vanishes in the vacuum at tree level? Consider first
the following bilinear (i.e. London type) Lagrangian:

$$
{\cal L}_L= -\frac{1}{4}(\partial_\mu A_{a\nu}-\partial_\nu A_{a\mu})^2 -
\frac{1}{8}(\epsilon^{\mu\nu\alpha\beta}\partial_\nu \Phi_ {a\alpha\beta})
(\epsilon_{\mu\eta\sigma\tau}\partial^\eta \Phi_ a^{\sigma\tau})
$$

\begin{equation}
\label{london}-\frac 14f_0^2\Phi _{a\mu \nu }\Phi _{a\mu \nu }+\frac
12g_0\Phi _{a\mu \nu }(\partial _\mu A_{a\nu }-\partial _\nu A_{a\mu })
\end{equation}
which possesses Abelianized gauge invariance $A_{a\mu }\rightarrow A_{a\mu
}+\partial _\mu \alpha _a$. The Lagrangian (\ref{london}) is intended to
form the basis for a truly interacting Ginzburg-Landau field system. The
order parameter is provided by the antisymmetric tensor $\Phi _{a\mu \nu
}=-\Phi _{a\nu \mu }$. It describes positive energy spin-1 excitations \cite
{GassLeut} with mass $f_0$ which correspond to three independent propagating
space-space components $\Phi _{amn}$. The constants $f_0^2\geq g_0^2$
characterize the ordered (confining) phase.

The polarization tensor $\Pi _{\mu \nu }\left( k\right) \equiv \left( k_\mu
k_\nu -k^2g_{\mu \nu }\right) \Pi \left( k^2\right) $ of the gluon field
relates the bare gluon propagator $D_{\mu \nu }\left( k\right) \equiv \left(
-g_{\mu \nu }+k_\mu k_\nu /k^2\right) /k^2$ in Landau gauge to the full
propagator $\Delta _{\mu \nu }$, in terms of the dielectric function $%
\epsilon \left( k^2\right) $:
\begin{equation}
\Delta _{\mu \nu }=\frac{D_{\mu \nu }}{1+\Pi \left( k^2\right) }\equiv \frac{%
D_{\mu \nu }}{\epsilon \left( k^2\right) }
\end{equation}

The propagator of the $\Phi $ field, deduced from (\ref{london}) is:
\begin{equation}
\label{phiprop}D_{\mu \nu ;\rho \sigma }(k)=\frac 1{k^2-f_0^2}(G_{\mu \nu
;\rho \sigma }-\frac 1{f_0^2}K_{\mu \nu ;\rho \sigma })
\end{equation}
where $G_{\mu \nu ;\rho \sigma }\equiv g_{\mu \rho }g_{\nu \sigma }-g_{\mu
\sigma }g_{\nu \rho }$ and $K_{\mu \nu ;\rho \sigma }\equiv k_\mu k_\rho
g_{\nu \sigma }-k_\nu k_\rho g_{\mu \sigma }-k_\mu k_\sigma g_{\nu \rho
}+k_\nu k_\sigma g_{\mu \rho }$. We can use the propagator eq.(\ref{phiprop}%
) and the interaction term in (\ref{london}) to derive thus the following
expression for the polarization tensor, in the tree approximation:
\begin{equation}
\Pi _{\nu \sigma }\left( k\right) =-g_0^2k^\mu k^\rho D_{\mu \nu ;\rho
\sigma }\left( k\right)
\end{equation}
It follows that $\Pi \left( k^2\right) =-g_0^2/f_0^2$ so that the dielectric
function is equal to:
\begin{equation}
\epsilon \left( k^2\right) =\frac{f_0^2-g_0^2}{f_0^2}
\end{equation}
We see that if $f_0=g_0$, the dielectric function vanishes and the
corresponding gluon propagator becomes infinite. By definition \cite{Chodos}%
, the simple Lagrangian (\ref{london}) describes the medium outside the
hadronic bags. Due to Lorentz invariance, which implies $\epsilon \mu =1$,
where $\mu $ is the magnetic permeability, the vacuum behaves simultaneously
as a perfect paramagnet.

In fact, when $f_0=g_0$, the Lagrangian (\ref{london}) can be rewritten in
the form \cite{ancestors}:

\begin{equation}
\label{ramond}{\cal L}_L=-\frac 18(\epsilon ^{\mu \nu \alpha \beta }\partial
_\nu \Phi _{a\alpha \beta })(\epsilon _{\mu \eta \sigma \tau }\partial ^\eta
\Phi _a^{\sigma \tau })-\frac 14(\partial _\mu A_{a\nu }-\partial _\nu
A_{a\mu }-g_0\Phi _{a\mu \nu })^2
\end{equation}
It may be checked that the Lagrangian (\ref{ramond}) depends only on the
field $(\partial _\mu A_{a\nu }-\partial _\nu A_{a\mu }-g_0\Phi _{a\mu \nu
}) $ so that the gluon field is effectively absorbed by the $\Phi $ field.
The Lagrangian (\ref{ramond}) exhibits a new type of gauge invariance:

\begin{equation}
\label{newgauge}\Phi _{a\mu \nu }\rightarrow \Phi _{a\mu \nu }+\partial _\mu
\xi _{a\nu }-\partial _\nu \xi _{a\mu }\quad \quad \quad A_{a\mu
}\rightarrow A_{a\mu }+g_0\xi _{a\mu }
\end{equation}
besides the original (Abelianized) gauge invariance $A_{a\mu }\rightarrow
A_{a\mu }+\partial _\mu \alpha _a$. Consequently, the massless gauge field $%
A_{a\mu }$ may be viewed as simply modifying the gauge of the $\Phi $ field.

If $g_0=0$ in the London Lagrangian (\ref{london}) the gauge field $A_{a\mu
} $ decouples from the massive order parameter $\Phi _{a\mu \nu }$ and the
dielectric function $\epsilon =1$. This is the regime we expect to prevail
inside a hadronic bag.

To achieve dynamically the transition between the vacuum phase $g_0=f_0$
outside the bag and the phase $g_0=0$ inside, we introduce a real colorless
scalar field $\chi (x)$ which develops a condensate in the vacuum due to a
potential
\begin{equation}
V(\chi )=\frac 12m^2\chi ^2-\frac 13m\xi \chi ^3+\frac 14\lambda \chi
^4\quad \quad \quad \left( m,\xi ,\lambda >0\right)
\end{equation}
For $\xi ^2>9\lambda /2$ the potential $V(\chi )$ has the essential feature
of presenting a minimum at a nonzero value $\chi _0\equiv M$:
\begin{equation}
\chi _0\equiv M=\frac 12m\frac \xi \lambda \left( 1+\sqrt{1-4\lambda /\xi ^2}%
\right)
\end{equation}
The coupling between the scalar field $\chi $ and the color fields is
achieved by replacing the constants $f_0$ and $g_0$ respectively by the
functions $f(\chi )$ and $g(\chi )$ which we choose to be the lowest order
polynomials $g^2(\chi )=(\alpha -1)\chi ^2$ and $f^2(\chi )=(\chi
-M)^2+(\alpha -1)\chi ^2$. This Ginzburg-Landau type extension of the London
Lagrangian (\ref{london}) is phenomenological and must be confronted to
experiment. Renormalizability, unlike in the Higgs case, is not a guiding
principle in such an effective low momentum theory. Thus an $SU(3)_c$ gauge
invariant Ginzburg-Landau type Lagrangian which reduces to (\ref{london})
for small $\Phi _a^{\mu \nu }$ and $A_a^\mu $ and for $\chi \rightarrow M$
is:

$$
{\cal L}_{GL} = - \frac{1}{4}F_{a\mu\nu}F_a^{\mu\nu} - \frac{1}{8}%
(\epsilon^{\mu\nu\alpha\beta}D_\nu \Phi_ {a\alpha\beta})
(\epsilon_{\mu\eta\sigma\tau}D^\eta \Phi_ a^{\sigma\tau})
$$

\begin{equation}
\label{nonabelian}-\frac 14f^2(\chi )\Phi _{a\mu \nu }\Phi _a^{\mu \nu
}+\frac 12g(\chi )\Phi _{a\mu \nu }F_a^{\mu \nu }+\frac 12(\partial _\mu
\chi )(\partial ^\mu \chi )-V(\chi )
\end{equation}

Two limiting cases corresponding to a constant $\chi $ field are of
interest. First, for a small gauge field, the equilibrium $\chi $ is
determined by solving $V^{\prime }\left( \chi \right) =0$ and the equation
of motion of the $\Phi _{a\mu \nu }$ field may be used to eliminate it in
favor of the gluon field, yielding:
\begin{equation}
\label{phif}\Phi _{a\mu \nu }=\frac{g(\chi )}{f^2(\chi )}F_{a\mu \nu }
\end{equation}
The Abelian version of the Lagrangian (\ref{nonabelian}) reduces then to
\cite{Kogut}:
\begin{equation}
\label{eff}{\cal L}_{eff}=-\frac 14(1-\frac{g^2(\chi )}{f^2(\chi )})F_{a\mu
\nu }F_a^{\mu \nu }+\frac 12(\partial _\mu \chi )(\partial ^\mu \chi
)-V(\chi )
\end{equation}
and the dielectric constant becomes:

\begin{equation}
\label{dielec}\epsilon (\chi )=\frac{f^2(\chi )-g^2(\chi )}{f^2(\chi )}=
\frac{(\chi -M)^2}{(\chi -M)^2+(\alpha -1)\chi ^2}
\end{equation}
It vanishes, as it should, in the vacuum ordered phase $\chi =M$.

Second, for a strong chromoelectric field, the equilibrium value of $\chi $
is determined by solving \cite{Lee}:
\begin{equation}
V_{eff}^{\prime }\left( \chi ,E_c\right) =\frac d{d\chi }\left( V\left( \chi
\right) +\frac{g^2\left( \chi \right) }{f^2\left( \chi \right) }\frac
12E_c^2\right) =0
\end{equation}
The term added to $V(\chi )$ increases with $E_c$. For a strong enough
chromoelectric field $E_c>E_c^{crit}$, the minimum of the effective
potential $V_{eff}(\chi ,E_c)$ will shift to $\chi =0$ thereby destroying
the ordered phase.

At finite temperatures, the Ginzburg-Landau theory is not a standard
finite-temperature theory. The use of the Lagrangian (\ref{nonabelian}) at
finite temperature requires an understanding of how its phenomenological
parameters depend on temperature. Notice that if $M$ decreases with
temperature there should appear a temperature$T_c$ at which the condensate $%
\chi $ disappears and the system returns then to the perturbative phase with
$\epsilon =1$.

We emphasize that the behavior of the gluon field in the presence of the
condensate $\chi =M$, which does not break any symmetry, is characteristic
of neither the perturbative nor the Higgs phase. Referring to t'Hooft's
phrasing \cite{t'Hooft} we believe that the Lagrangian (\ref{nonabelian})
with the condensate $\chi =M$ is an example of a realization of the color
confined phase.

\medskip

\begin{center}
{\Large {\bf Acknowledgements}}
\end{center}

This work was done during a stay of J.H. at the Service de Physique
Th\'eorique de Saclay supported by the ECC grant 3317. The authors thank
J.P.Blaizot and J.Stern for most valuable advice.


\begin{thebibliography}{9}
\bibitem{thirring}  J.Hosek, Phys.Lett. {\bf 226B, }377 (1989); H.Narnhofer
and W.Thirring, in Springer Tracts Modern Physics {\bf 119, }1 (1990).

\bibitem{Lee}  T.D.Lee, Particle Physics and Introduction to Field theory,
Harwood Academic, New-York, 1981.

\bibitem{Kogut}  J.Kogut and L.Susskind, Phys.Rev. {\bf D9}, 3501 (1974).

\bibitem{Chodos}  A.Chodos et al. Phys.Rev. {\bf D9}, 3471 (1974); for a
review see P.Hasenfrantz and J.Kuti, Phys.Rep. {\bf 40C}, 75 (1978).

\bibitem{GassLeut}  J.Gasser and H.Leutwyler, Annals Phys. {\bf 158}, 142
(1984).

\bibitem{ancestors}  The Lagrangian eq.(\ref{ramond}) has been considered in
various contexts; M.Kalb and P.Ramond, Phys.Rev.{\bf D9}, 2273 (1974);
E.Cremmer and J.Scherk, Nucl.Phys {\bf B72}, 117 (1974); Y.Nambu, Phys.Rep.
{\bf 23C}, 250 (1976); M.Baker, J.S.Ball and F.Zachariasen, Phys.Rep. {\bf %
209C}, 73 (1991); H.Kleinert, Phys.Lett. {\bf B293}, 168 (1992).

\bibitem{t'Hooft}  G.t'Hooft, Nucl.Phys. {\bf B138}, 1 (1978); {\bf B153},
141 (1979).
\end{thebibliography}
\end{document}